\documentclass[twocolumn,aps]{revtex4-1}
\usepackage{amsmath}
\usepackage{braket}
\usepackage{cancel}
\usepackage{mathtools} 
\usepackage{mathbbol}
\usepackage{siunitx}
\usepackage{qcircuit}
\usepackage[pdftex]{graphicx}
\usepackage{amsfonts}
\usepackage{amssymb}
\usepackage{braket}
\usepackage{siunitx}
\usepackage{graphicx}
\usepackage{tikz} 
\usepackage{calc}
\usepackage{float}
\usepackage{color}
\usepackage{mathtools}

\newcommand{\inter}{\Phi}

\begin{document} 
\title{Mitigated barren plateaus in the time-nonlocal optimization of analog quantum-algorithm protocols}
\author{Lukas Broers$^{1,2}$ and Ludwig Mathey$^{1,2,3}$}
\affiliation{
$^{1}$Center for Optical Quantum Technologies, University of Hamburg, 22761 Hamburg, Germany\\
$^{2}$Institute for Quantum Physics, University of Hamburg, 22761 Hamburg, Germany\\
$^{3}$The Hamburg Center for Ultrafast Imaging, 22761 Hamburg, Germany}
\begin{abstract}
Quantum machine learning has emerged as a promising utilization of near-term quantum computation devices.
However, algorithmic classes such as variational quantum algorithms have been shown to suffer from barren plateaus due to vanishing gradients in their parameters spaces.
We present an approach to quantum algorithm optimization that is based on trainable Fourier coefficients of Hamiltonian system parameters.
Our ansatz is exclusive to the extension of discrete quantum variational algorithms to analog quantum optimal control schemes and is non-local in time.
We demonstrate the viability of our ansatz on the objectives of compiling the quantum Fourier transform and preparing ground states of random problem Hamiltonians.
In comparison to the temporally local discretization ansätze in quantum optimal control and parameterized circuits, our ansatz exhibits faster and more consistent convergence.
We uniformly sample objective gradients across the parameter space and find that in our ansatz the variance decays at a non-exponential rate with the number of qubits, while it decays at an exponential rate in the temporally local benchmark ansatz.
This indicates the mitigation of barren plateaus in our ansatz.
We propose our ansatz as a viable candidate for near-term quantum machine learning. 
\end{abstract}
\maketitle
 
\section{Introduction} 

Quantum machine learning (QML) connects classical machine learning and quantum information processing. 
This emergent field promises new methods that advance quantum computation \cite{qmlrev, QMLchallenges} and has brought forth a class of approaches referred to as variational quantum algorithms (VQA) \cite{VQAreview,VQAprog,VQES,bonetmonroig23}. 
In particular, noisy intermediate-scale quantum (NISQ) devices \cite{bharti2021noisy,Preskill2018quantumcomputingin,McClean_2016} are predicted to benefit from the synergies with machine learning found in VQA. 
These approaches optimize parameters in a sequence of unitary operations, the product of which describes the time-evolution of the system. 
The optimization is performed with respect to a chosen observable.
Examples include quantum approximate optimization algorithms (QAOA) \cite{QAOA,QAOA3}, quantum neural networks \cite{questforqnn,qpower,DLQNN,sharma2020trainability,QCNN,pesah2020absence,Cerezo_2021}, quantum circuit learning \cite{QCL} and quantum assisted quantum-compiling \cite{QAQC,QAOAuniversal2,UVQA}. 

\begin{figure}
\centering 
\includegraphics[width=0.84\linewidth]{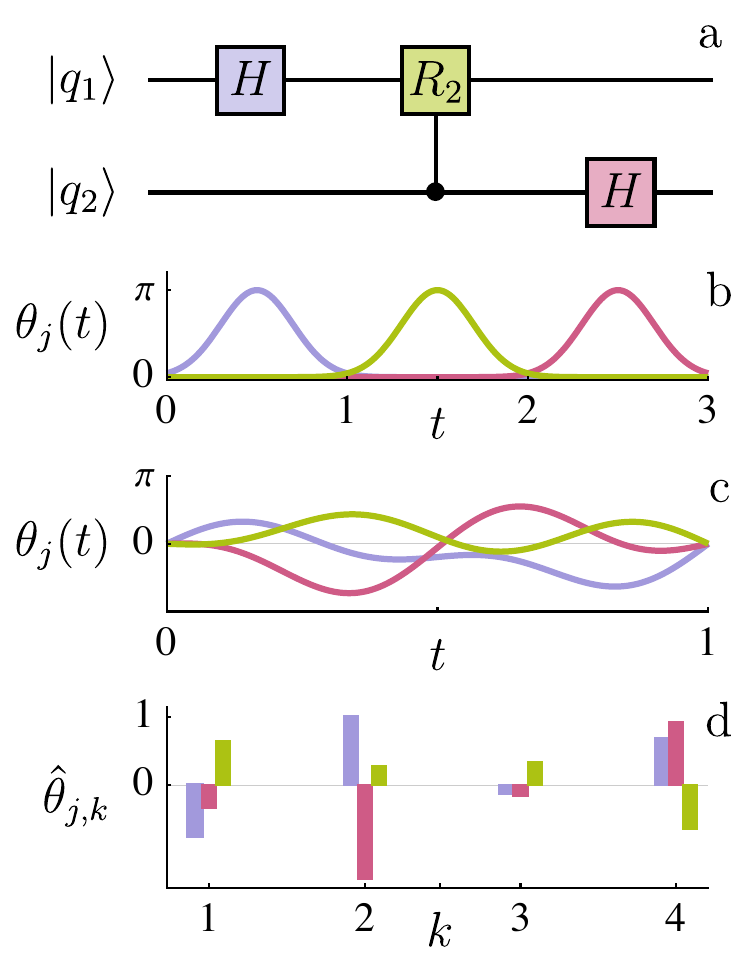}
\caption{
{\bf Levels of abstraction of quantum algorithms}.
A common formulation of quantum circuits consists of a set of discrete gates, see panel (a).
The physical realization of these gates consists of temporally isolated control protocols of the system parameters.
These are denoted as $\theta_j(t)$ for the different parameters, see panel (b).
A more efficient realization utilizes the full space of temporal evolutions of the parameters $\theta_j(t)$.
This includes fully parallel protocols which take less time to complete the task, see panel (c).
Any such protocol can be expressed via its Fourier coefficients $\hat{\theta}_{j,k}$, which we specifically treat as trainable parameters in our ansatz, see panel (d).
}\label{fig1}
\end{figure} 
  
Similarly, quantum optimal control (QOC) aims to optimize the time-dependent system parameters of a quantum system to attain a given objective \cite{CRAB,CRAB2,lloyd14,qocrev,lijun17,GOAT,qocreview}.
QOC has been connected to VQA approaches, and advantages of moving from the discrete circuit picture to the underlying physical system parameters have been demonstrated \cite{Magann, QOCVQA}.
Such analog VQA approaches commonly utilize piecewise constant, or step-wise, parameterization ansätze \cite{stepwiseOC,RLQOC,RLgates,Pontryagin,principle}, which behave like the Trotterized limit of very deep parameterized quantum circuits with very small actions per gate.

A major obstacle of VQA is the existence of barren plateaus in the loss landscapes, i.e. increasingly large regimes in the parameter space with exponentially vanishing gradients, which hinder training \cite{NIBPs,entanglement1,entanglement2,barrenplateaus,diagbarren,arrasmith2021equivalence,Volkoff_2021,Anschuetz22}.
The general scaling behavior and emergence of barren plateaus is largely not understood and the dependence of barren plateaus on the details of VQA has been an active field of research in recent years.
The comparison of local to global objective functions, the dependence on circuit depth, and the effects of spatial and temporal locality of parameterizations have been studied in connection to barren plateaus \cite{barrenplateaus,localcost,arrasmith2021equivalence,Uvarov_2021,diagbarren,localtdesign}.
In particular, the emergence of barren plateaus has been proven in time-locally parameterized quantum circuits for global objective functions and for local objective functions in the case of non-shallow circuits \cite{localcost,barrenplateaus,arrasmith2021equivalence}.
Limiting the controllability of such ansätze can reduce the onset of barren plateaus \cite{holmes2021connecting,localtdesign,anschuetz2023critical,perezsalinas2023reducechop}, which constitutes a trade-off in expressibility \cite{rabitz,solomon} in favor of trainability.
This includes ansätze that are tailored to a given problem, such as the variational Hamiltonian ansatz \cite{VQAprog,park2023hamiltonian}, the unitary coupled cluster ansatz \cite{UCC19}, and QAOA \cite{QAOA3}. 
These results suggest that non-local ansätze that depart from the parameterized circuit paradigm may mitigate barren plateaus without the loss of generality.
Overcoming the obstacle of barren plateaus is crucial for the success of near-term QML technologies.

In this paper, we propose a parameterization ansatz for quantum algorithm optimization using generalized analog protocols.
In this ansatz we directly control the Fourier coefficients of the system parameters of a Hamiltonian.
This constitutes a method that is non-local in time and is exclusive to analog quantum protocols as it does not translate into discrete circuit parameterizations which are conventionally found in VQA. 
We compare our ansatz to the common optimal control ansatz of step-wise parameterizations for the example objectives of compiling the quantum Fourier transform as well as minimizing the energy of random problem Hamiltonians.
This comparison shows that this Fourier based ansatz results in solutions with higher fidelity and in particular superior convergence behavior.
Note that the optimization of Fourier coefficients has been proposed for the control of molecular dynamics \cite{kormann10}.
It has also been used in a mixed approach that optimizes in the basis of piecewise constant functions \cite{songyao22}, as well as in phase-modulated gradient-free optimization \cite{jiazhao20}.
The Fourier basis has also been used with tuned frequencies in the CRAB algorithm \cite{CRAB, Scheuer14}.
However, studies on this particular ansatz in the context of analog quantum computing as a natural extension of VQA appear to be lacking.
We demonstrate that our ansatz exhibits non-exponential scaling behavior with respect to the number of qubits in the objective gradient variance, which suggests the absence of barren plateaus.
We conclude that our ansatz is a promising candidate for efficient training and avoiding barren plateaus in VQA.

\section{Methods}

In quantum circuits, the time-dependent Hamiltonian parameters that implement the gates are sequential, rather than parallel, and therefore contain long idling times. 
This is a consequence of deconstructing unitary transformations into algorithmic sequences of logical gates.
Fig.~\ref{fig1} illustrates different levels of abstraction of quantum algorithms. 
The departure from the conventional quantum circuit paradigm towards a larger and more intricate space of solutions of quantum protocols enables a computational speed-up due to parallelized Hamiltonian operations.

We write a general time-dependent Hamiltonian as 
\begin{equation}
H(t) = \sum_{j} \theta_j(t) H_j,
\label{genham}
\end{equation}
where $H_j$ are Hermitian matrices that define the system.
$\theta_j(t)$ are the parameters that determine the time-dependence of the system.
The resulting time-evolution operator is formally written as
\begin{equation}
    U_{\boldsymbol\theta} = \hat{T}[\exp\{ -i\int_0^1 \sum_{j} \theta_j(t) H_j dt \}],
    \label{uthetaearly}
\end{equation}
where $\hat{T}$ indicates time-ordering. 
We restrict the time-evolution to $t\in[0,1]$ and use units in which $\hbar=1$, for simplicity.
The unitary transformation $U_{\boldsymbol\theta}$ is explicitly a function of the protocols $\theta_j(t)$. 
In order to perform numerical optimization, it is necessary to choose a particular parameterization for the $\theta_j(t)$.  

In the ansatz which we highlight in this work, we parameterize the $\theta_j(t)$ in terms of the first $n_\mathrm{f}$ real-valued Fourier coefficients $\theta_{j,k}$ such that
\begin{equation}
\theta_j(t) = \sum_{k=1}^{n_{\mathrm{f}}} {\theta}_{j,k} \sin(\pi k t).
\label{hamfourier}
\end{equation}
This ansatz is motivated by its inherent temporal non-locality, as varying a single parameter $\theta_{j,k}$ changes the protocol $\theta_j(t)$ at all points in time. 
It presents a natural choice for a time-non-local parameterization that results in protocols that are smooth and slowly varying by construction, which is advantageous experimentally.
We initialize the parameters $\theta_{j,k}$ randomly between $\pm \pi/k$, such that slow modes are emphasized.

In addition to our ansatz, we consider the step-wise ansatz that uses the common discretization in terms of piecewise constant system parameters.
\begin{equation}
    \theta_j(t) = {\theta}_{j,k}, \frac{k}{n_{\mathrm{f}}} \leq t < \frac{k+1}{n_{\mathrm{f}}},
    \label{hamstep}
\end{equation}
with $k=0,\dots,n_\mathrm{f}-1$.  
We initialize the ${\theta}_{j,k}$ randomly between $\pm\pi$. 
This ansatz is time-local and generates discontinuous step-functions with $n_f$ steps with values $\theta_{j,k}$.
These steps are reminiscent of the sequences of parameterized gates in quantum circuits as they are conventionally found in VQA.
Due to its connection to conventional parameterized variational circuits, this ansatz serves as a benchmark to which we compare our ansatz of Eq.~\ref{hamfourier}.

In either ansatz, we optimize the parameters 
\begin{equation} 
    \boldsymbol\theta = \sum_{j,k}\theta_{j,k}\hat{e}_{j,k},
\end{equation}
with respect to a given objective function $\mathcal{L}_{\boldsymbol\theta}$, which encodes a target transformation.
The exact expression of any objective $\mathcal{L}_\mathbf{\theta}$ depends on the details of the problem it describes.
The $\hat{e}_{j,k}$ are formally constructed unit-vectors that collect the trainable parameters $\theta_{j,k}$ in the vector $\boldsymbol\theta$.
Successful optimization corresponds to a time-evolution operator $U_{\boldsymbol\theta}$ which implements the target transformation.
For a single optimization iteration, we vary the individual parameters $\theta_{j,k}$ by a small $\delta$ and evaluate the objective function to estimate the respective derivatives 
\begin{equation}
\frac{\partial \mathcal{L}_{\boldsymbol\theta}}{\partial \theta_{j,k}} \approx \frac{\mathcal{L}_{\boldsymbol\theta+\delta \hat{e}_{j,k}} - \mathcal{L}_{\boldsymbol\theta}}{\delta}
\end{equation}
such that we obtain the gradient 
\begin{equation}
\nabla \mathcal{L}_{\boldsymbol\theta} = \sum_{j,k} \frac{\partial \mathcal{L}_{\boldsymbol\theta}}{\partial \theta_{j,k}}.
\end{equation}
We then update the parameters as 
\begin{equation}
\boldsymbol\theta_\mathrm{old} \rightarrow \boldsymbol\theta_\mathrm{new} =\boldsymbol\theta_\mathrm{old} - \eta g^\dagger \nabla \mathcal{L}_{\boldsymbol{\theta}},
\end{equation}
where $\eta$ is the learning rate, which we update dynamically using the ADAM \cite{adam} algorithm.
$g$ is the Fubini-Study metric, which contains information on the quantum geometry of the system in order to improve training behavior and makes this approach a quantum natural gradient descent method \cite{QNG}.
For more details see App.~\ref{app:qng}.

Note that in a physical realization, the parameters $\theta_j(t)$ cannot become arbitrarily large, and are limited by physical constraints or features of the realization.
In our numerical approach, these parameters are unbounded. 
However, we find that these parameters remain reasonably small throughout learning, as we show below.

\begin{figure}
\centering 
\includegraphics[width=1.0\linewidth]{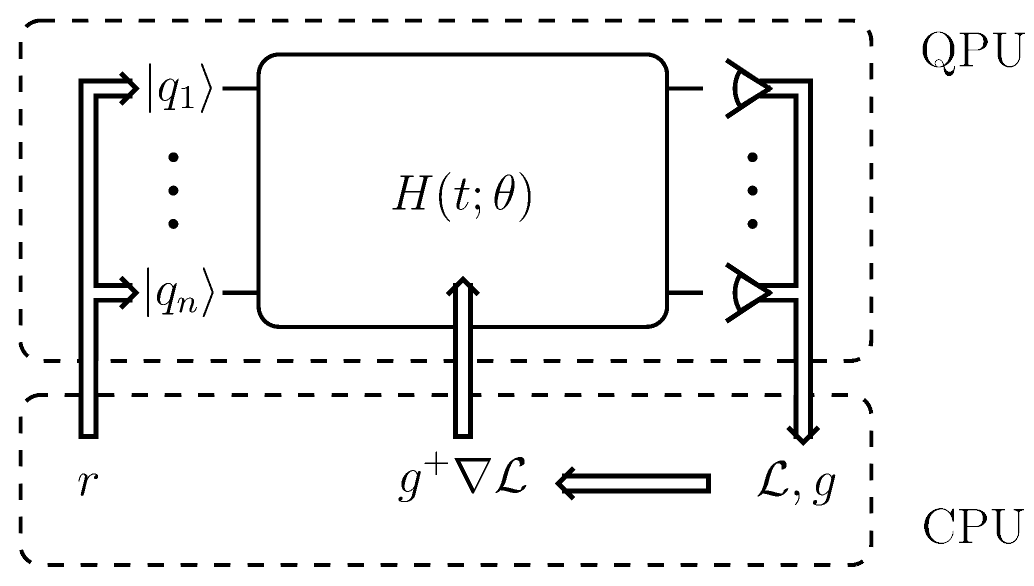}
\caption{
{\bf Illustration of hybrid quantum optimization}.
A quantum processing unit (QPU) is assumed to have controllable parameters ${\theta}$.
Problem-specific input $r$ is mapped onto the initial state of the qubits which the QPU evolves in time according to the parameters $\theta$ and its underlying Hamiltonian $H$.
The final qubit state is measured to determine the value of an objective function $\mathcal{L}_{\boldsymbol\theta}$ and the Fubini-Study metric $g$.
These quantities are used on a classical machine to approximate the quantum natural gradient step to update the parameters $\theta$ and improve $\mathcal{L}_{\boldsymbol\theta}$.
}\label{fig2}
\end{figure}

\section{Results}
We compare our Fourier ansatz to the step-wise ansatz for the objectives of quantum compiling and energy minimization.
Further, we evaluate the scaling behavior of the variances of objective gradients with respect the number of qubits.
Throughout this work we use the Ising Hamiltonian \cite{ising73} with a two-component transverse field for $n_\mathrm{q}$ qubits as the controllable system that generates the variational unitary $U_{\boldsymbol\theta}$.
It is
\begin{equation}
H(t) = \sum_{j=1}^{n_\mathrm{q}} \left(B^j_x(t)\sigma_x^j + B^j_y(t) \sigma_y^j\right) + \sum_{j=1}^{n_\mathrm{q}-1}J_j (t)\sigma_z^{j}\sigma_z^{j+1},
\label{exham}
\end{equation}
with controllable parameters $B^j_x(t)$, $B^j_y(t)$ and $J_j(t)$.
We consider open boundary conditions, such that the index of $J_j(t)$ goes up to $j=n_\mathrm{q}-1$.
In total this gives $(3n_\mathrm{q}-1) n_{\mathrm{f}}$ trainable parameters in $\boldsymbol{\theta}$, as the $B_x^j(t)$, $B_y^j(t)$ and $J_j(t)$ take the role of the $\theta_j(t)$ in Eq.~\ref{genham}.
Our ansatz In Eq.~\ref{hamfourier} presents a general parameterization of system parameters and therefore the particular choice of the Hamiltonian is not essential. 
In particular, neither the Fourier ansatz nor the choice of the Hamiltonian are informed \textit{a priori} by any objective at hand. 
They are agnostic to the optimizational tasks we utilize them for.

\subsection{Quantum Compiling}
We first demonstrate the performance of our ansatz for the example of learning implementations of the QFT represented by the unitary operation $V$, operating on $n_\mathrm{q}$ qubits. 
The matrix elements of $V$ are 
\begin{equation}
V_{k,l} = 2^{-\frac{n_\mathrm{q}}{2}} \exp\{i 2\pi k l 2^{-n_\mathrm{q}}\},
\label{qft}
\end{equation}
where $k,l=1,\dots,2^{n_\mathrm{q}}$.
For compiling unitary transformations, we utilize the objective function 
\begin{equation}
    \mathcal{L}^U_{\boldsymbol\theta}= 1 - \frac{1}{|\{r\}|}\sum_r |\braket{r|U^\dagger_{\boldsymbol\theta} V|r}|^2,
    \label{unitaryloss}
\end{equation}
where $\{r\}$ is a set of randomized unentangled input states
\begin{equation}
\ket{r} = \otimes_{i=1}^{n_\mathrm{q}}[\cos(\frac{\phi_i}{2})\ket{0}+e^{i \psi_i}\sin(\frac{\phi_i}{2})\ket{1}],
\end{equation}  
which is similar to recent methods \cite{OODdynamics}.
This objective function estimates the implementation error $\epsilon=1-|\mathrm{Tr}(U^\dagger_\theta V)2^{-n_\mathrm{q}}|^2$ between the unitaries $U_{\boldsymbol\theta}$ and $V$.
Note that there exist state estimation and tomography methods \cite{classicalshadows,nntomography,permuttomography,efficienttomography,variationaltomography} that are experimentally favorable over the overlap in Eq.~\ref{unitaryloss}.
Here we use this overlap due to its straightforward numerical accessibility.

\begin{figure}
\centering 
\includegraphics[width=1.0\linewidth]{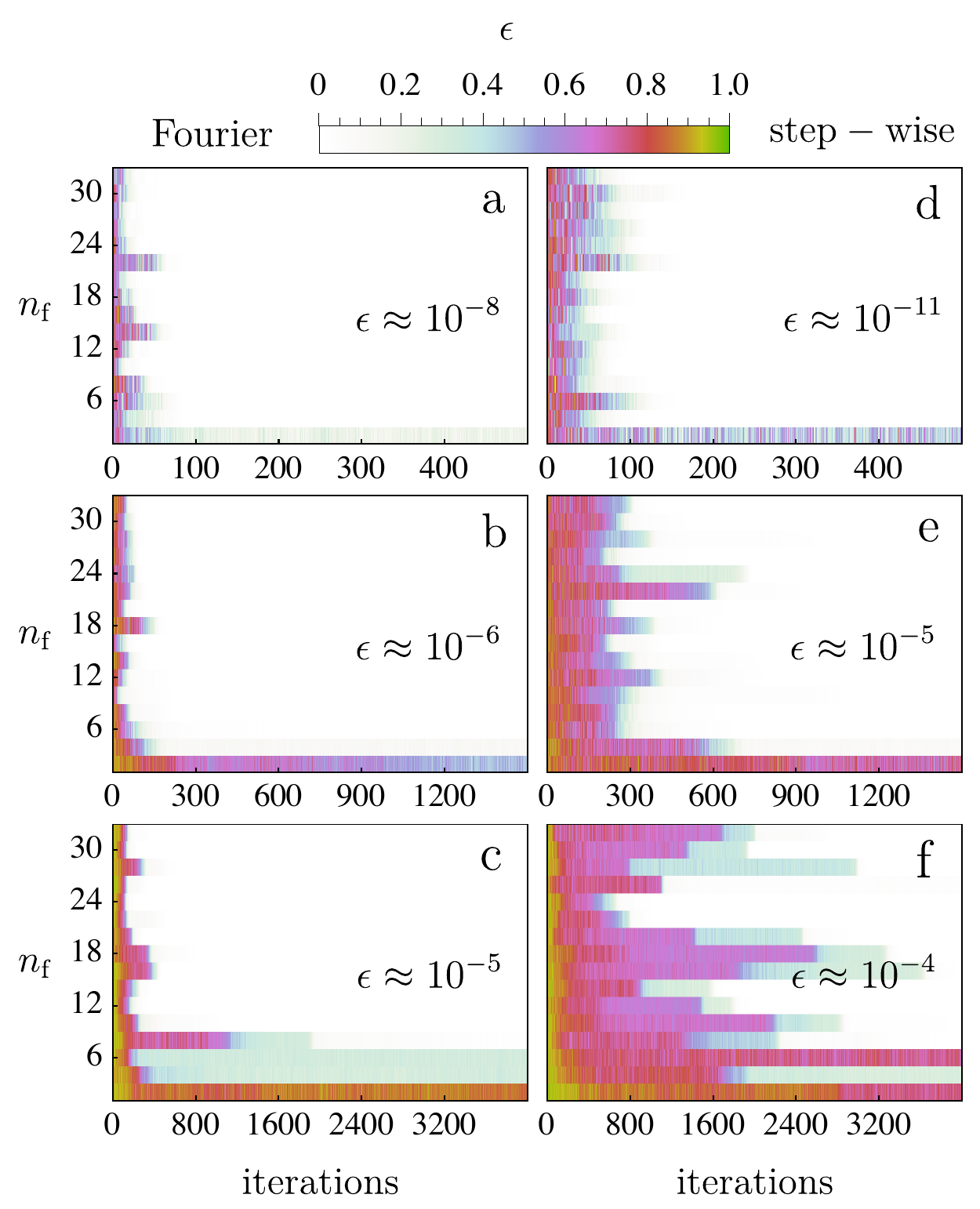}
\caption{
{\bf Implementation errors during training of the quantum Fourier transform}.
The errors $\epsilon$ during training as a function of the hyperparameter $n_{\mathrm{f}}$ for the QFT for $n_\mathrm{q}=2$ (a,d), $3$ (b,e) and $4$ (c,f) for our Fourier based ansatz (a,b,c) and the step-wise protocol ansatz (d,e,f).
For sufficiently large $n_{\mathrm{f}} \geq n_{\mathrm{f},\mathrm{min}}$ both ansätze converge to very small errors.
Our Fourier based ansatz outperforms the step-wise ansatz in terms of convergence speed and consistency.
}\label{fig3}
\end{figure}

In Fig.~\ref{fig3}, we show the estimated implementation error $\epsilon$ during training, as a function of $n_\mathrm{f}$ for $n_\mathrm{q}\leq4$.
We observe that both implementations converge to the target transformation for sufficiently large $n_{\mathrm{f}}$. 
For smaller $n_{\mathrm{f}}$ the accessible unitary transformations generated from the ansätze Eqs.~\ref{hamfourier} and \ref{hamstep} are insufficient and presumably do not contain the QFT on $n_\mathrm{q}$ qubits.
 
We emphasize that our Fourier based ansatz is consistently outperforming the step-wise ansatz in terms of convergence speed. 
We show in Figs.~\ref{fig3}~(a,b,c) that our ansatz tends to converge after roughly $50$, $100$ and $200$ training iterations for $n_\mathrm{q}=2$, $3$ and $4$, respectively.
Figs.~\ref{fig3}~(d,e,f) show that the step-wise protocol ansatz tends to converge after roughly $100$, $300$ and $1800$ iterations for $n_\mathrm{q}=2,3,4$, respectively.
For $n_\mathrm{q}=4$ in Fig.~\ref{fig3}~(f), the convergence behavior of the step-wise ansatz is increasingly inconsistent.
The step-wise ansatz has the tendency to linger at suboptimal fidelities from which it only moves away very slowly. 
This behavior becomes more prominent with increasing $n_\mathrm{q}$ and is a consequence of the loss landscape that follows from the parameterization in Eq.~\ref{hamstep}.
Our ansatz does not show this behavior, but rather exhibits faster and more direct convergence. 
This is an indication for the absence of vanishing gradients, as is apparent when comparing Figs.~\ref{fig3}~(c) and (f).

\begin{figure}
\centering
\includegraphics[width=1.0\linewidth]{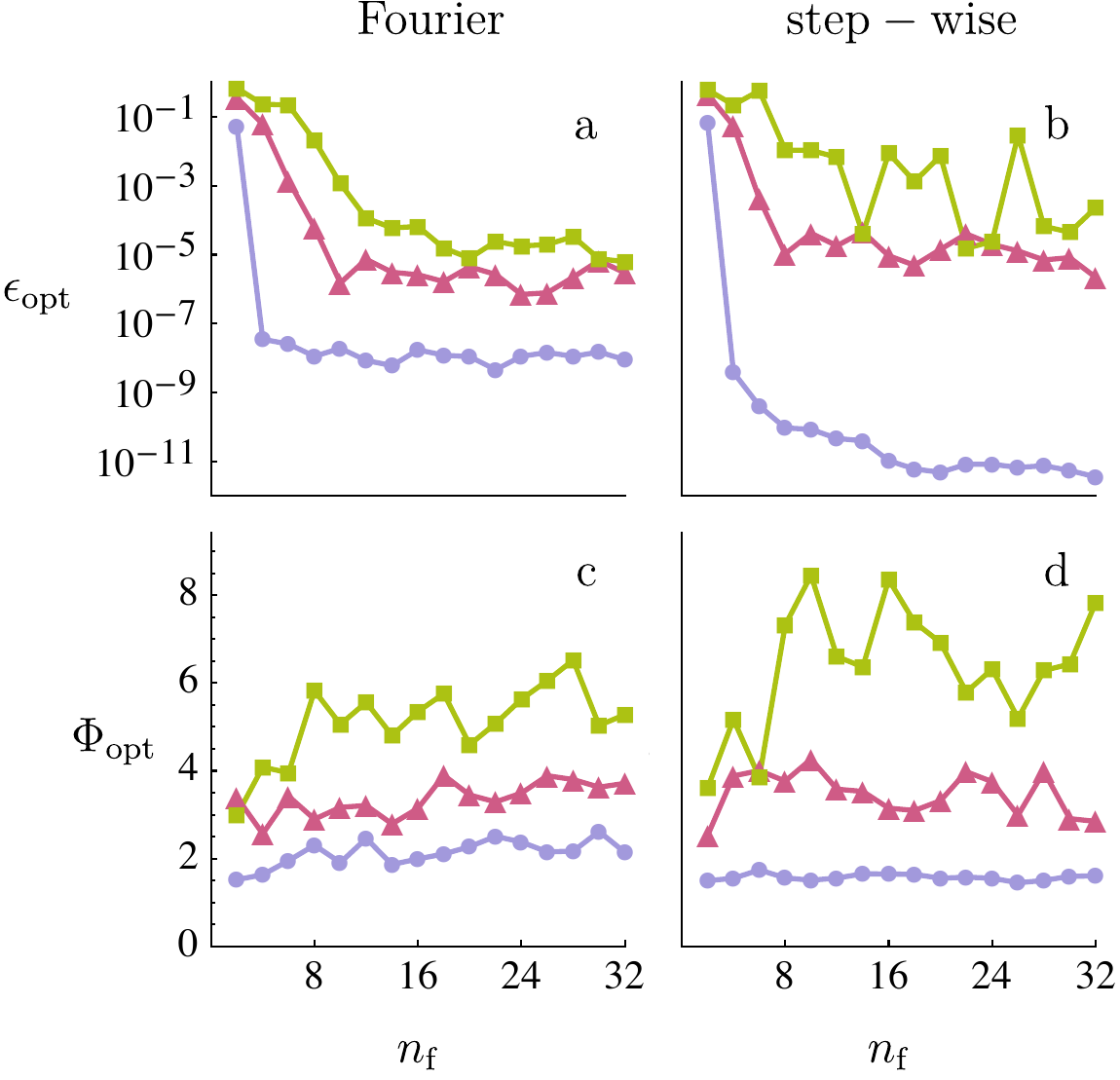}
\caption{
{\bf Minimal errors and effective actions for training the quantum Fourier transform}.
The minimal errors $\epsilon_\mathrm{opt}$ (a,b) found during training and the corresponding effective protocol actions $\inter_\mathrm{opt}$ (c,d) of both ansätze. 
The training results are for the QFT for $n_\mathrm{q}=2$ (blue circles), $3$ (red triangles) and $4$ (green squares).
The inconsistent $\epsilon_\mathrm{opt}$ in the step-wise ansatz for $n_\mathrm{q}=4$ (b) is a consequence of the suboptimal convergence behavior, related to the emergence of barren plateaus.
}\label{fig4}
\end{figure}

In order to further evaluate the quality of the converged solutions, we show the minimal errors after training $\epsilon_\mathrm{opt}$ with respect to the hyperparameter $n_{\mathrm{f}}$ for both ansätze in Figs.~\ref{fig4}~(a) and (b).
We find the minimal $n_{\mathrm{f}}$ that is necessary for convergence during training to be approximately $n_{\mathrm{f},\mathrm{min}}\approx 4$, $6$ and $8$ for $n_\mathrm{q}=2$, $3$ and $4$, respectively.
The minimal $n_{\mathrm{f}}$ necessary for convergence appears to be the same for both ansätze in this example. 
For larger $n_{\mathrm{f}}$, the minimal error converges to very small values that show no strong dependence on $n_{\mathrm{f}}$. 
For the cases of $n_\mathrm{q}=3$ and $n_\mathrm{q}=4$, the resulting minimal error tends to approach $\epsilon_\mathrm{opt}\approx 10^{-5}$.  
We note that for a concrete experimental realization, additional considerations, e.g. what dissipative processes are present and how well a specific parameter can be tuned dynamically, determine the overall success of these approaches, which will be explored elsewhere.

As a second figure of merit we consider the effective implementation action, which we quantify with the integrated magnitude of the vector of system parameters ${{\boldsymbol\theta}}(t)$, such that
\begin{equation}
\inter = \int_0^1  |{{\boldsymbol\theta}}(t)| \mathrm{d}t.
\end{equation}
Given that the parameters ${\theta}_j(t)$ have the units of energy, this quantity is an overall measure of the phase or action that is accumulated during the time-evolution.
It therefore quantifies an estimate of both the energy that is required to implement a protocol in a given time, as well as the time that is required given a bound to the magnitude of the parameters $\theta_j(t)$. 
This figure of merit allows us to determine whether a solution with improved fidelity in our Fourier ansatz merely emerges due to decreased time-efficiency.
In Figs.~\ref{fig4}~(c) and (d) we show the effective actions $\inter_\mathrm{opt}$ of the same optimal solutions of Figs.~\ref{fig4}~(a) and (b), with respect to the hyperparameter $n_{\mathrm{f}}$.
We find the two ansätze to be very similar in terms of necessary action and therefore time-efficiency.
In both ansätze, there is no strong dependence on the hyperparameter $n_{\mathrm{f}}$ past $n_{\mathrm{f},\mathrm{min}}$. 
While the implementation actions consistently remain reasonably small, there is a clear and expected tendency of implementations to require larger effective actions with increasing amounts of qubits.

\subsection{Energy Minimization}
\begin{figure}
\centering 
\includegraphics[width=1.0\linewidth]{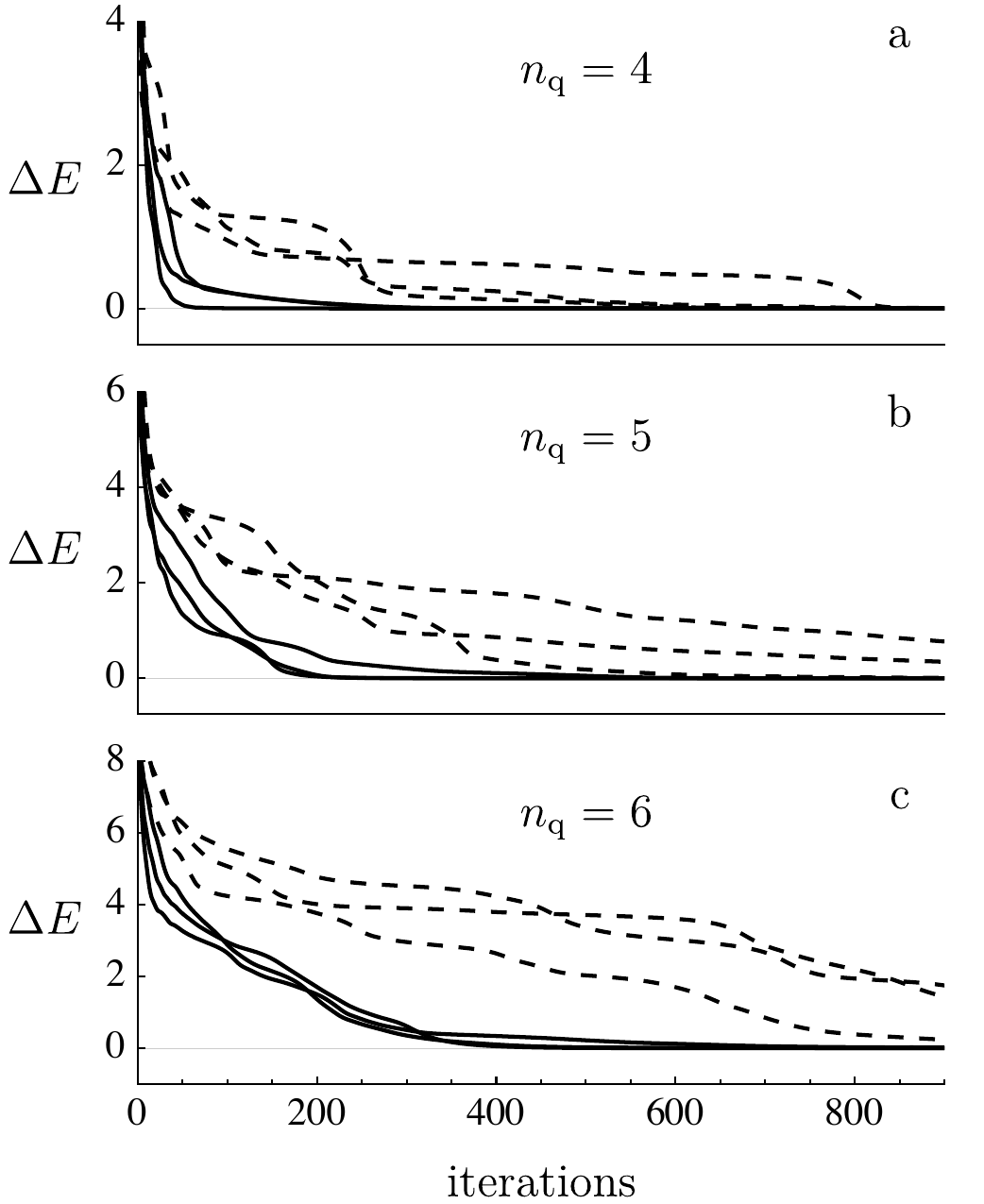}
\caption{
{\bf Training trajectories for energy minimization}.
Learning trajectories for the ground state preparation of three randomly generated problem Hamiltonians for $n_\mathrm{q}=4,5,6$ qubits for our ansatz (solid lines) and the step-wise ansatz (dashed lines).
$\Delta E=\braket{E}_{\boldsymbol\theta}-E_0$ is the expected energy of the prepared states relative to the ground state energy. In all cases $n_{\mathrm{f}}=16$.}\label{fig5}
\end{figure}

As a second optimization task, we consider the energy expectation value of a problem Hamiltonian $H_{\mathrm{p}}$ and its minimization.
Specifically, we consider the objective function
\begin{equation}
    \mathcal{L}^E_{\boldsymbol\theta}=\braket{E}_{\boldsymbol\theta} =\braket{0|U^\dagger_{\boldsymbol\theta} H_{\mathrm{p}} U_{\boldsymbol\theta}|0},
    \label{energymin}
\end{equation}
where $U_{\boldsymbol\theta}$ is the time-evolution operator of the Hamiltonian given in Eq.~\ref{uthetaearly}, which we use to construct the trial state $U_{\boldsymbol\theta}\ket{0}$. 
We use the shortened notation $\ket{0}=\ket{0}^{\otimes n_\mathrm{q}}$ of the state where all qubits are in the logical zero state.
We perform this ground state search for random problem Hamiltonians for both our ansatz and the step-wise ansatz with $n_{\mathrm{f}}=16$.
In this example we do not apply the QNG, i.e. we set the metric $g=\mathbb{1}$, for simplicity.
Fig.~\ref{fig5} shows the energy differences to the ground state energies $\Delta E = \braket{E}_{\boldsymbol\theta} - E_0$ for the training trajectories of three randomized problem Hamiltonians for up to six qubits.
We again see that our ansatz outperforms the step-wise ansatz in terms of convergence speed.
There is an increasing tendency of gradients to flatten out in the step-wise ansatz.
This behavior is not present in our ansatz and indicates the onset of barren plateaus in the optimization of ground state preparation for step-wise protocols.

\subsection{Objective Gradient Variances}
In order to quantify the presence of barren plateaus, we consider the variance of the gradients of the objective function for both our ansatz and the step-wise ansatz.
In random parameterized quantum circuits this amounts to uniformly sampling possible initializations in the parameter space of $\boldsymbol{\theta}$ \cite{barrenplateaus}.
In analog parameterizations of quantum algorithms, the parameter space is aperiodic and non-compact such that sampling is more intricate.
We consider uniformly sampled vectors $\boldsymbol{\theta}$ inside a $(3n_\mathrm{q}-1)$-dimensional ball with radius $|\boldsymbol{\theta}|_\mathrm{max}$ for each time-step in the step-wise ansatz, and  $|\boldsymbol{\theta}|_\mathrm{max}/k$ for each $k$th Fourier mode in our ansatz.
The value of $|\boldsymbol{\theta}|_\mathrm{max}$ determines the set of reachable states of a given ansatz. 
We consider the variance of the gradient with respect to the first parameter 
\begin{equation}
\mathrm{Var}[\partial_{\theta_{1,1}} \mathcal{L}^E_{\boldsymbol\theta}]
=
\braket{(\partial_{\theta_{1,1}}\mathcal{L}^E_{\boldsymbol\theta})^2}-\braket{\partial_{\theta_{1,1}}\mathcal{L}^E_{\boldsymbol\theta}}^2
\end{equation}
for the specific problem Hamiltonian 
\begin{equation}
    H_\mathrm{p} = \sigma_z^{1}\sigma_z^{2}\prod_{j= 3}^{n_\mathrm{q}}\mathbb{1}^{j}.
    \label{hp}
\end{equation}

We calculate the variance as a function of $|\boldsymbol{\theta}|_\mathrm{max}$ for up to 8 qubits for $n_\mathrm{f}=128$.
Analytical arguments on the existence of barren plateaus in RPQCs \cite{barrenplateaus} rely on time-local expressions of the gradient of a loss function such as Eq.~\ref{energymin}.
This also applies to the step-wise ansatz. 
However, in our ansatz given by Eq.~\ref{hamfourier}, the expression is
\begin{equation} 
    \partial_{\boldsymbol{\theta}_{1,1}}\mathcal{L}^E_{\boldsymbol{\theta}} = i \int_0^1 \sin(\pi t') \braket{0|U^0_{t'} [U^{t'}_1 H_\mathrm{p} U^1_{t'},H_1] U_{0}^{t'} |0} dt',
\end{equation} 
where $U_a^b$ is the time-evolution operator from the time $a$ to the time $b \geq a$. For $a \geq b$ it is $U_a^b={(U^a_b)}^\dagger$.
The variance of this expression includes all possible covariances of time-local changes to the protocols $\boldsymbol{\theta}(t)$, which differs substantially from the variances in RPQCs.
Further, in the parameter space of $\boldsymbol{\theta}(t)$, the unitaries $U_0^{t'}$ and $U_{t'}^1$ are neither necessarily independent in the sense of the Haar measure nor guaranteed to be 2-designs.
Therefore, the analytical argument for RPQCs \cite{barrenplateaus} does not apply to our ansatz.
In particular, the argument generates no statement about the scaling behavior.

\begin{figure}
\centering 
\includegraphics[width=1.0\linewidth]{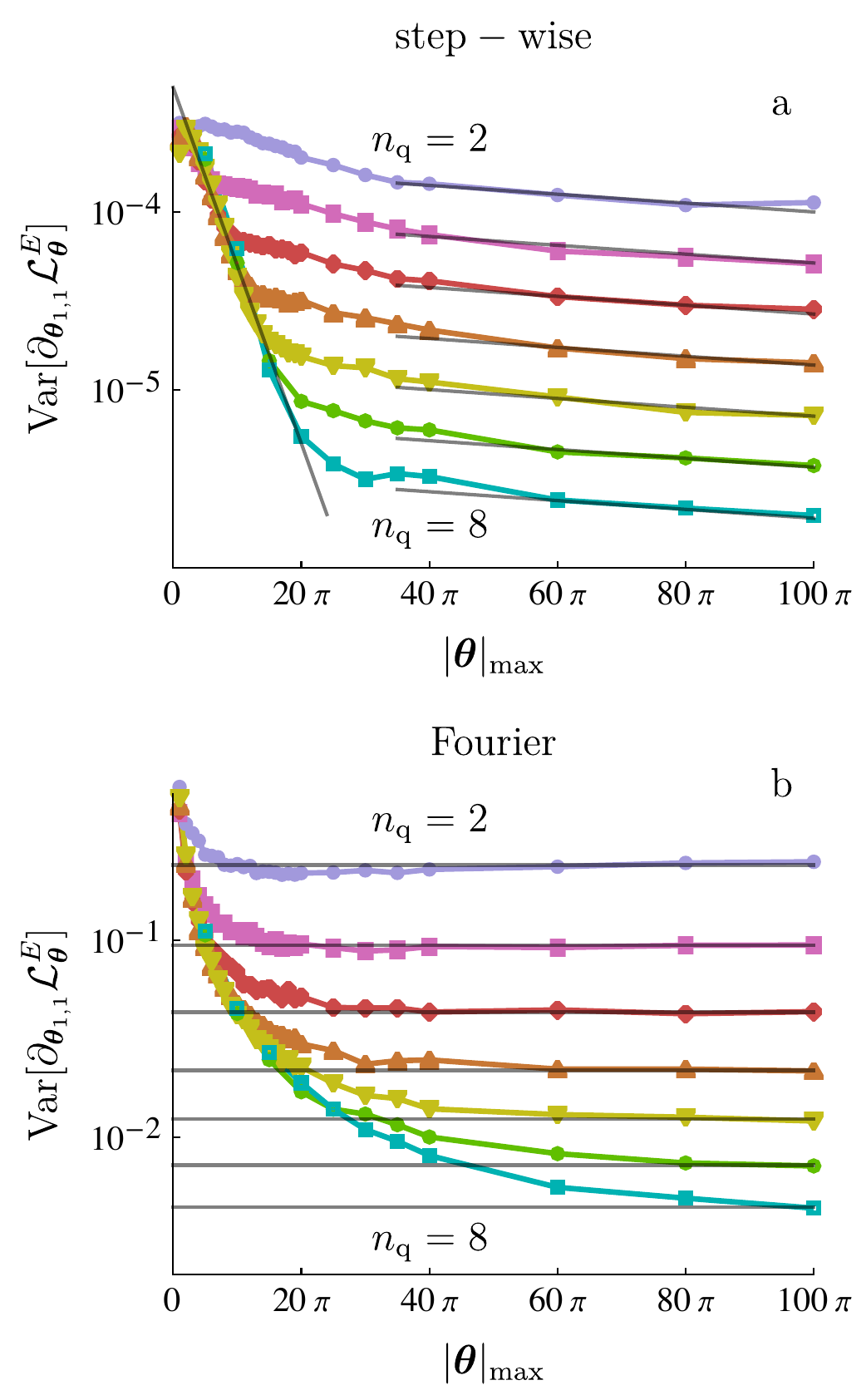}
\caption{
{\bf Variances of the energy objective gradient. }
The variance of the gradient $\partial_{{\boldsymbol\theta}_{1,1}}\mathcal{L}_{\boldsymbol{\theta}}^E$ of the loss function $\mathcal{L}^E_{\boldsymbol \theta}=\braket{0|U_{\boldsymbol\theta}^\dagger [\sigma_z^{1} \sigma_z^{2}] U_{\boldsymbol\theta}|0}$ for up to $8$ qubits for the step-wise ansatz (a) and our Fourier based ansatz (b) on a logarithmic scale.
The parameters are sampled uniformly within a radius of $|\boldsymbol{\theta}|_\mathrm{max}$ for $n_\mathrm{f}=128$. 
The lines are visual guides. 
}\label{fig6}
\end{figure}

In Fig.~\ref{fig6}~(a), we show the results of the step-wise ansatz.
We find that the variance is independent of the amount of qubits $n_\mathrm{q}$ for small $|\boldsymbol{\theta}|_\mathrm{max}$.
For increasing $|\boldsymbol{\theta}|_\mathrm{max}$, the variance decays exponentially with $|\boldsymbol{\theta}|_\mathrm{max}$ with slopes that are independent of $n_\mathrm{q}$.
More importantly, the variance decays exponentially as a function of $n_\mathrm{q}$ with a log-scale slope of roughly $\ln(\frac{1}{2})$, as indicated by the equally spaced lines.
The step-wise ansatz is reminiscent of a continuous Trotterized limit of parameterized circuits and therefore these results are consistent with barren plateau studies on RPQCs \cite{barrenplateaus}.

In Fig.~\ref{fig6}~(b), we show the results for our ansatz. 
The variances show asymptotic behavior as functions of $|\boldsymbol{\theta}|_\mathrm{max}$.
They converge at increasingly large values of $|\boldsymbol{\theta}|_\mathrm{max}$, which vastly exceed implementation actions that are necessary for highly entangling unitaries such as the QFT as we show in Fig.~\ref{fig4}~(c).
Thus, in our ansatz $|\boldsymbol{\theta}|_\mathrm{max}$ provides a useful hyperparameter for initialization that can be tuned to comparatively small values where the scaling with $n_\mathrm{q}$ is very favorable.
Further, we find that the variance decreases as a function of $n_\mathrm{q}$ at a decreasing and non-exponential rate.
This non-exponential scaling behavior indicates the reduction of barren plateaus in our ansatz, in particular during initialization.

\section{Conclusion}
  
We have proposed a system-agnostic ansatz of analog variational quantum algorithms rooted in quantum optimal control.
The central feature of our ansatz is that it treats the Fourier coefficients of the time-controlled system parameters of a given Hamiltonian as trainable. 
Therefore, our ansatz is non-local in time and has no direct analog in discretized parameterized quantum circuits. 
By restricting the modes to low-end frequencies we keep the amount of trainable parameters low, while also ensuring smooth quantum protocols and sufficient controllability by construction.
We have applied a measurement based stochastic quantum natural gradient optimization scheme to our ansatz to generate protocols for the quantum Fourier transform for up to four qubits.
Additionally, we have optimized ground state preparation processes for random problem Hamiltonians for up to six qubits.
We compared the results to optimizations of the more commonly utilized step-wise parameterization ansatz.
The results we have presented here are limited to few-qubit systems, as the numerical simulations on the native Hamiltonian level are computationally more demanding than the circuit-based counter-parts of conventional VQA. 
This does not translate into a lack of scalability in a true hybrid realization of the proposed method.

We have demonstrated that the convergence behavior of our ansatz outperforms the step-wise protocols in speed and consistency.
We have found the effective implementation action to be comparable and to remain reasonably small in both ansätze.
We have analyzed the gradient along the loss landscape for both ansätze, and have shown that our ansatz shows non-exponentially decreasing variances with respect to the amount of qubits, indicating an absence of barren plateaus.
The step-wise ansatz shows a characteristic exponential decay with the amount of qubits that is consistent with barren plateau studies on random parameterized quantum circuits.
The scaling behavior of objective gradient variances for larger systems, as well as tuning the sampling range for initialization and its relation to expressibility, will be elaborated on elsewhere.

In conclusion, our ansatz is a promising candidate for mitigating barren plateaus in quantum algorithm optimization and presents an alternative to parameterizations that are discrete or local in time.
This approach is of direct relevance for current efforts of implementing quantum computing, as it provides realistic and efficient access to optimal quantum algorithm protocols.

\acknowledgements{
This work is funded by the Deutsche Forschungsgemeinschaft (DFG, German Research Foundation) -- SFB-925 -- project 170620586,
and the Cluster of Excellence 'Advanced Imaging of Matter' (EXC 2056), Project No. 390715994.
}

\bibliographystyle{unsrt}
\bibliography{lit}

\begin{thebibliography}{10}

\bibitem{qmlrev}
Jacob Biamonte, Peter Wittek, Nicola Pancotti, Patrick Rebentrost, Nathan
  Wiebe, and Seth Lloyd.
\newblock Quantum machine learning.
\newblock {\em Nature}, 549(7671):195--202, 2017.

\bibitem{QMLchallenges}
M.~Cerezo, Guillaume Verdon, Hsin-Yuan Huang, Lukasz Cincio, and Patrick~J.
  Coles.
\newblock Challenges and opportunities in quantum machine learning.
\newblock {\em Nature Computational Science}, 2(9):567--576, 2022.

\bibitem{VQAreview}
M.~Cerezo, Andrew Arrasmith, Ryan Babbush, Simon~C. Benjamin, Suguru Endo,
  Keisuke Fujii, Jarrod~R. McClean, Kosuke Mitarai, Xiao Yuan, Lukasz Cincio,
  and Patrick~J. Coles.
\newblock Variational quantum algorithms.
\newblock {\em Nature Reviews Physics}, 3(9):625--644, 2021.

\bibitem{VQAprog}
Dave Wecker, Matthew~B. Hastings, and Matthias Troyer.
\newblock Progress towards practical quantum variational algorithms.
\newblock {\em Phys. Rev. A}, 92:042303, Oct 2015.

\bibitem{VQES}
Alberto Peruzzo, Jarrod McClean, Peter Shadbolt, Man-Hong Yung, Xiao-Qi Zhou,
  Peter~J. Love, Al{\'a}n Aspuru-Guzik, and Jeremy~L. O'Brien.
\newblock A variational eigenvalue solver on a photonic quantum processor.
\newblock {\em Nature Communications}, 5(1):4213, 2014.

\bibitem{bonetmonroig23}
Xavier Bonet-Monroig, Hao Wang, Diederick Vermetten, Bruno Senjean, Charles
  Moussa, Thomas B\"ack, Vedran Dunjko, and Thomas~E. O'Brien.
\newblock Performance comparison of optimization methods on variational quantum
  algorithms.
\newblock {\em Phys. Rev. A}, 107:032407, Mar 2023.

\bibitem{bharti2021noisy}
Kishor Bharti, Alba Cervera-Lierta, Thi~Ha Kyaw, Tobias Haug, Sumner
  Alperin-Lea, Abhinav Anand, Matthias Degroote, Hermanni Heimonen, Jakob~S.
  Kottmann, Tim Menke, Wai-Keong Mok, Sukin Sim, Leong-Chuan Kwek, and Al\'an
  Aspuru-Guzik.
\newblock Noisy intermediate-scale quantum algorithms.
\newblock {\em Rev. Mod. Phys.}, 94:015004, Feb 2022.

\bibitem{Preskill2018quantumcomputingin}
John Preskill.
\newblock Quantum {C}omputing in the {NISQ} era and beyond.
\newblock {\em {Quantum}}, 2:79, August 2018.

\bibitem{McClean_2016}
Jarrod~R McClean, Jonathan Romero, Ryan Babbush, and Al{\'{a}}n Aspuru-Guzik.
\newblock The theory of variational hybrid quantum-classical algorithms.
\newblock {\em New Journal of Physics}, 18(2):023023, feb 2016.

\bibitem{QAOA}
Leo Zhou, Sheng-Tao Wang, Soonwon Choi, Hannes Pichler, and Mikhail~D. Lukin.
\newblock Quantum approximate optimization algorithm: Performance, mechanism,
  and implementation on near-term devices.
\newblock {\em Phys. Rev. X}, 10:021067, Jun 2020.

\bibitem{QAOA3}
Stuart Hadfield, Zhihui Wang, Bryan O’Gorman, Eleanor~G. Rieffel, Davide
  Venturelli, and Rupak Biswas.
\newblock From the quantum approximate optimization algorithm to a quantum
  alternating operator ansatz.
\newblock {\em Algorithms}, 12(2), 2019.

\bibitem{questforqnn}
Maria Schuld, Ilya Sinayskiy, and Francesco Petruccione.
\newblock The quest for a quantum neural network.
\newblock {\em Quantum Information Processing}, 13(11):2567--2586, 2014.

\bibitem{qpower}
Amira Abbas, David Sutter, Christa Zoufal, Aurelien Lucchi, Alessio Figalli,
  and Stefan Woerner.
\newblock The power of quantum neural networks.
\newblock {\em Nature Computational Science}, 1(6):403--409, 2021.

\bibitem{DLQNN}
Kerstin Beer, Dmytro Bondarenko, Terry Farrelly, Tobias~J. Osborne, Robert
  Salzmann, Daniel Scheiermann, and Ramona Wolf.
\newblock Training deep quantum neural networks.
\newblock {\em Nature Communications}, 11(1):808, 2020.

\bibitem{sharma2020trainability}
Kunal Sharma, M.~Cerezo, Lukasz Cincio, and Patrick~J. Coles.
\newblock Trainability of dissipative perceptron-based quantum neural networks.
\newblock {\em Phys. Rev. Lett.}, 128:180505, May 2022.

\bibitem{QCNN}
Iris Cong, Soonwon Choi, and Mikhail~D. Lukin.
\newblock Quantum convolutional neural networks.
\newblock {\em Nature Physics}, 15(12):1273--1278, 2019.

\bibitem{pesah2020absence}
Arthur Pesah, M.~Cerezo, Samson Wang, Tyler Volkoff, Andrew~T. Sornborger, and
  Patrick~J. Coles.
\newblock Absence of barren plateaus in quantum convolutional neural networks.
\newblock {\em Phys. Rev. X}, 11:041011, Oct 2021.

\bibitem{Cerezo_2021}
M~Cerezo and Patrick~J Coles.
\newblock Higher order derivatives of quantum neural networks with barren
  plateaus.
\newblock {\em Quantum Science and Technology}, 6(3):035006, jun 2021.

\bibitem{QCL}
K.~Mitarai, M.~Negoro, M.~Kitagawa, and K.~Fujii.
\newblock Quantum circuit learning.
\newblock {\em Phys. Rev. A}, 98:032309, Sep 2018.

\bibitem{QAQC}
Sumeet Khatri, Ryan LaRose, Alexander Poremba, Lukasz Cincio, Andrew~T.
  Sornborger, and Patrick~J. Coles.
\newblock Quantum-assisted quantum compiling.
\newblock {\em {Quantum}}, 3:140, May 2019.

\bibitem{QAOAuniversal2}
M.~E.~S. Morales, J.~D. Biamonte, and Z.~Zimbor{\'a}s.
\newblock On the universality of the quantum approximate optimization
  algorithm.
\newblock {\em Quantum Information Processing}, 19(9):291, 2020.

\bibitem{UVQA}
Jacob Biamonte.
\newblock Universal variational quantum computation.
\newblock {\em Phys. Rev. A}, 103:L030401, Mar 2021.

\bibitem{CRAB}
Patrick Doria, Tommaso Calarco, and Simone Montangero.
\newblock Optimal control technique for many-body quantum dynamics.
\newblock {\em Phys. Rev. Lett.}, 106:190501, May 2011.

\bibitem{CRAB2}
Tommaso Caneva, Tommaso Calarco, and Simone Montangero.
\newblock Chopped random-basis quantum optimization.
\newblock {\em Phys. Rev. A}, 84:022326, Aug 2011.

\bibitem{lloyd14}
S.~Lloyd and S.~Montangero.
\newblock Information theoretical analysis of quantum optimal control.
\newblock {\em Phys. Rev. Lett.}, 113:010502, Jul 2014.

\bibitem{qocrev}
Steffen~J. Glaser, Ugo Boscain, Tommaso Calarco, Christiane~P. Koch, Walter
  K{\"o}ckenberger, Ronnie Kosloff, Ilya Kuprov, Burkhard Luy, Sophie Schirmer,
  Thomas Schulte-Herbr{\"u}ggen, Dominique Sugny, and Frank~K. Wilhelm.
\newblock Training schr{\"o}dinger's cat: quantum optimal control.
\newblock {\em The European Physical Journal D}, 69(12):279, 2015.

\bibitem{lijun17}
Jun Li, Xiaodong Yang, Xinhua Peng, and Chang-Pu Sun.
\newblock Hybrid quantum-classical approach to quantum optimal control.
\newblock {\em Phys. Rev. Lett.}, 118:150503, Apr 2017.

\bibitem{GOAT}
Shai Machnes, Elie Ass\'emat, David Tannor, and Frank~K. Wilhelm.
\newblock Tunable, flexible, and efficient optimization of control pulses for
  practical qubits.
\newblock {\em Phys. Rev. Lett.}, 120:150401, Apr 2018.

\bibitem{qocreview}
Christiane~P. Koch, Ugo Boscain, Tommaso Calarco, Gunther Dirr, Stefan Filipp,
  Steffen~J. Glaser, Ronnie Kosloff, Simone Montangero, Thomas
  Schulte-Herbr{\"u}ggen, Dominique Sugny, and Frank~K. Wilhelm.
\newblock Quantum optimal control in quantum technologies. strategic report on
  current status, visions and goals for research in europe.
\newblock {\em EPJ Quantum Technology}, 9(1):19, 2022.

\bibitem{Magann}
Alicia~B. Magann, Christian Arenz, Matthew~D. Grace, Tak-San Ho, Robert~L.
  Kosut, Jarrod~R. McClean, Herschel~A. Rabitz, and Mohan Sarovar.
\newblock From pulses to circuits and back again: A quantum optimal control
  perspective on variational quantum algorithms.
\newblock {\em PRX Quantum}, 2:010101, Jan 2021.

\bibitem{QOCVQA}
Alexandre Choquette, Agustin Di~Paolo, Panagiotis~Kl. Barkoutsos, David
  S\'en\'echal, Ivano Tavernelli, and Alexandre Blais.
\newblock Quantum-optimal-control-inspired ansatz for variational quantum
  algorithms.
\newblock {\em Phys. Rev. Research}, 3:023092, May 2021.

\bibitem{stepwiseOC}
Navin Khaneja, Timo Reiss, Cindie Kehlet, Thomas Schulte-Herbrüggen, and
  Steffen~J. Glaser.
\newblock Optimal control of coupled spin dynamics: design of nmr pulse
  sequences by gradient ascent algorithms.
\newblock {\em Journal of Magnetic Resonance}, 172(2):296--305, 2005.

\bibitem{RLQOC}
Marin Bukov, Alexandre G.~R. Day, Dries Sels, Phillip Weinberg, Anatoli
  Polkovnikov, and Pankaj Mehta.
\newblock Reinforcement learning in different phases of quantum control.
\newblock {\em Phys. Rev. X}, 8:031086, Sep 2018.

\bibitem{RLgates}
Murphy~Yuezhen Niu, Sergio Boixo, Vadim~N. Smelyanskiy, and Hartmut Neven.
\newblock Universal quantum control through deep reinforcement learning.
\newblock {\em npj Quantum Information}, 5(1):33, 2019.

\bibitem{Pontryagin}
Zhi-Cheng Yang, Armin Rahmani, Alireza Shabani, Hartmut Neven, and Claudio
  Chamon.
\newblock Optimizing variational quantum algorithms using pontryagin's minimum
  principle.
\newblock {\em Phys. Rev. X}, 7:021027, May 2017.

\bibitem{principle}
V.G. Boltyanski, R.V. Gamkrelidze, E.F. Mishchenko, and L.S. Pontryagin.
\newblock The maximum principle in the theory of optimal processes of control.
\newblock {\em IFAC Proceedings Volumes}, 1(1):464--469, 1960.
\newblock 1st International IFAC Congress on Automatic and Remote Control,
  Moscow, USSR, 1960.

\bibitem{NIBPs}
Samson Wang, Enrico Fontana, M.~Cerezo, Kunal Sharma, Akira Sone, Lukasz
  Cincio, and Patrick~J. Coles.
\newblock Noise-induced barren plateaus in variational quantum algorithms.
\newblock {\em Nature Communications}, 12(1):6961, 2021.

\bibitem{entanglement1}
Carlos Ortiz~Marrero, M\'aria Kieferov\'a, and Nathan Wiebe.
\newblock Entanglement-induced barren plateaus.
\newblock {\em PRX Quantum}, 2:040316, Oct 2021.

\bibitem{entanglement2}
Taylor~L. Patti, Khadijeh Najafi, Xun Gao, and Susanne~F. Yelin.
\newblock Entanglement devised barren plateau mitigation.
\newblock {\em Phys. Rev. Research}, 3:033090, Jul 2021.

\bibitem{barrenplateaus}
Jarrod~R. McClean, Sergio Boixo, Vadim~N. Smelyanskiy, Ryan Babbush, and
  Hartmut Neven.
\newblock Barren plateaus in quantum neural network training landscapes.
\newblock {\em Nature Communications}, 9(1):4812, 2018.

\bibitem{diagbarren}
Martin Larocca, Piotr Czarnik, Kunal Sharma, Gopikrishnan Muraleedharan,
  Patrick~J. Coles, and M.~Cerezo.
\newblock Diagnosing {B}arren {P}lateaus with {T}ools from {Q}uantum {O}ptimal
  {C}ontrol.
\newblock {\em {Quantum}}, 6:824, September 2022.

\bibitem{arrasmith2021equivalence}
Andrew Arrasmith, Zo{\"e} Holmes, M~Cerezo, and Patrick~J Coles.
\newblock Equivalence of quantum barren plateaus to cost concentration and
  narrow gorges.
\newblock {\em Quantum Science and Technology}, 7(4):045015, aug 2022.

\bibitem{Volkoff_2021}
Tyler Volkoff and Patrick~J Coles.
\newblock Large gradients via correlation in random parameterized quantum
  circuits.
\newblock {\em Quantum Science and Technology}, 6(2):025008, jan 2021.

\bibitem{Anschuetz22}
Eric~R. Anschuetz and Bobak~T. Kiani.
\newblock Quantum variational algorithms are swamped with traps.
\newblock {\em Nature Communications}, 13(1):7760, 2022.

\bibitem{localcost}
M.~Cerezo, Akira Sone, Tyler Volkoff, Lukasz Cincio, and Patrick~J. Coles.
\newblock Cost function dependent barren plateaus in shallow parametrized
  quantum circuits.
\newblock {\em Nature Communications}, 12(1):1791, 2021.

\bibitem{Uvarov_2021}
A~V Uvarov and J~D Biamonte.
\newblock On barren plateaus and cost function locality in variational quantum
  algorithms.
\newblock {\em Journal of Physics A: Mathematical and Theoretical},
  54(24):245301, May 2021.

\bibitem{localtdesign}
Fernando G. S.~L. Brand{\~a}o, Aram~W. Harrow, and Micha{\l} Horodecki.
\newblock Local random quantum circuits are approximate polynomial-designs.
\newblock {\em Communications in Mathematical Physics}, 346(2):397--434, 2016.

\bibitem{holmes2021connecting}
Zo\"e Holmes, Kunal Sharma, M.~Cerezo, and Patrick~J. Coles.
\newblock Connecting ansatz expressibility to gradient magnitudes and barren
  plateaus.
\newblock {\em PRX Quantum}, 3:010313, Jan 2022.

\bibitem{anschuetz2023critical}
Eric~R. Anschuetz.
\newblock Critical points in quantum generative models, 2023.

\bibitem{perezsalinas2023reducechop}
Adrián Pérez-Salinas, Radoica Draškić, Jordi Tura, and Vedran Dunjko.
\newblock Reduce\&chop: Shallow circuits for deeper problems, 2023.

\bibitem{rabitz}
Viswanath Ramakrishna and Herschel Rabitz.
\newblock Relation between quantum computing and quantum controllability.
\newblock {\em Phys. Rev. A}, 54:1715--1716, Aug 1996.

\bibitem{solomon}
S.~G. Schirmer, H.~Fu, and A.~I. Solomon.
\newblock Complete controllability of quantum systems.
\newblock {\em Phys. Rev. A}, 63:063410, May 2001.

\bibitem{park2023hamiltonian}
Chae-Yeun Park and Nathan Killoran.
\newblock Hamiltonian variational ansatz without barren plateaus, 2023.

\bibitem{UCC19}
Joonho Lee, William~J. Huggins, Martin Head-Gordon, and K.~Birgitta Whaley.
\newblock Generalized unitary coupled cluster wave functions for quantum
  computation.
\newblock {\em Journal of Chemical Theory and Computation}, 15(1):311--324, 01
  2019.

\bibitem{kormann10}
Katharina Kormann, Sverker Holmgren, and Hans~O. Karlsson.
\newblock A fourier-coefficient based solution of an optimal control problem in
  quantum chemistry.
\newblock {\em Journal of Optimization Theory and Applications},
  147(3):491--506, 2010.

\bibitem{songyao22}
Yao Song, Junning Li, Yong-Ju Hai, Qihao Guo, and Xiu-Hao Deng.
\newblock Optimizing quantum control pulses with complex constraints and few
  variables through autodifferentiation.
\newblock {\em Phys. Rev. A}, 105:012616, Jan 2022.

\bibitem{jiazhao20}
Jiazhao Tian, Haibin Liu, Yu~Liu, Pengcheng Yang, Ralf Betzholz, Ressa~S. Said,
  Fedor Jelezko, and Jianming Cai.
\newblock Quantum optimal control using phase-modulated driving fields.
\newblock {\em Phys. Rev. A}, 102:043707, Oct 2020.

\bibitem{Scheuer14}
Jochen Scheuer, Xi~Kong, Ressa~S Said, Jeson Chen, Andrea Kurz, Luca Marseglia,
  Jiangfeng Du, Philip~R Hemmer, Simone Montangero, Tommaso Calarco, Boris
  Naydenov, and Fedor Jelezko.
\newblock Precise qubit control beyond the rotating wave approximation.
\newblock {\em New Journal of Physics}, 16(9):093022, sep 2014.

\bibitem{adam}
Diederik~P. Kingma and Jimmy Ba.
\newblock Adam: A method for stochastic optimization.
\newblock {\em CoRR}, abs/1412.6980, 2015.

\bibitem{QNG}
James Stokes, Josh Izaac, Nathan Killoran, and Giuseppe Carleo.
\newblock Quantum {N}atural {G}radient.
\newblock {\em {Quantum}}, 4:269, May 2020.

\bibitem{ising73}
R~B Stinchcombe.
\newblock Ising model in a transverse field. i. basic theory.
\newblock {\em Journal of Physics C: Solid State Physics}, 6(15):2459, aug
  1973.

\bibitem{OODdynamics}
Matthias~C. Caro, Hsin-Yuan Huang, Nicholas Ezzell, Joe Gibbs, Andrew~T.
  Sornborger, Lukasz Cincio, Patrick~J. Coles, and Zo{\"e} Holmes.
\newblock Out-of-distribution generalization for learning quantum dynamics.
\newblock {\em Nature Communications}, 14(1):3751, 2023.

\bibitem{classicalshadows}
Ryan Levy, Di~Luo, and Bryan~K. Clark.
\newblock Classical shadows for quantum process tomography on near-term quantum
  computers, 2021.

\bibitem{nntomography}
Marcel Neugebauer, Laurin Fischer, Alexander J\"ager, Stefanie Czischek, Selim
  Jochim, Matthias Weidem\"uller, and Martin G\"arttner.
\newblock Neural-network quantum state tomography in a two-qubit experiment.
\newblock {\em Phys. Rev. A}, 102:042604, Oct 2020.

\bibitem{permuttomography}
G.~T\'oth, W.~Wieczorek, D.~Gross, R.~Krischek, C.~Schwemmer, and
  H.~Weinfurter.
\newblock Permutationally invariant quantum tomography.
\newblock {\em Phys. Rev. Lett.}, 105:250403, Dec 2010.

\bibitem{efficienttomography}
Marcus Cramer, Martin~B. Plenio, Steven~T. Flammia, Rolando Somma, David Gross,
  Stephen~D. Bartlett, Olivier Landon-Cardinal, David Poulin, and Yi-Kai Liu.
\newblock Efficient quantum state tomography.
\newblock {\em Nature Communications}, 1(1):149, 2010.

\bibitem{variationaltomography}
Ryan LaRose, Arkin Tikku, {\'E}tude O'Neel-Judy, Lukasz Cincio, and Patrick~J.
  Coles.
\newblock Variational quantum state diagonalization.
\newblock {\em npj Quantum Information}, 5(1):57, 2019.

\end{thebibliography}
 
\appendix
 
\section{Quantum Natural Gradient}
\label{app:qng}
In order to estimate the gradient of $\mathcal{L}_{\boldsymbol\theta}$, we modify a single component ${\theta}_{j,k}$ by a small amount $\delta=10^{-7}$. 
This results in slightly altered time-evolution operators $U^{j,k}_{\boldsymbol\theta} = U_{{\boldsymbol\theta}+\delta \hat{e}_{j,k}}$ and values for the objective $\mathcal{L}_{{\boldsymbol\theta}+\delta \hat{e}_{j,k}}$.
This gives access to the finite difference estimate
\begin{equation}
\frac{\partial \mathcal{L_{\boldsymbol\theta}}}{\partial {\theta}_{j,k}} \approx \frac{\mathcal{L}_{{\boldsymbol\theta}+\delta \hat{e}_{j,k} }-\mathcal{L}_{{\boldsymbol\theta}}}{\delta}.
\end{equation}
We do this for all possible $j$ and $k$ and write
\begin{equation}
\vec{\nabla} \mathcal{L}_{\boldsymbol\theta}  = \sum_{j,k} \frac{\partial \mathcal{L_{\boldsymbol\theta}}}{\partial {\theta}_{j,k}}\hat{e}_{j,k}.
\end{equation}
The quantum natural gradient update $\Delta{\boldsymbol\theta}$ is then given by \cite{QNG}
\begin{equation}
g(\Delta {\boldsymbol\theta}) = - \eta \vec\nabla \mathcal{L}_{{\boldsymbol\theta}} 
\label{qng} 
\end{equation}
where $\eta$ is a dynamical learning rate following the ADAM algorithm with standard parameters and a step-size of $0.01$ \cite{adam}. 
The quantum natural gradient considers the underlying geometry of the parameterized states using the Fubini-Study metric $g$ which has the components
\begin{align}
g_{(i,k)}^{(j,l)} &= \mathrm{Re}[\braket{\partial_{{\boldsymbol\theta}_{i,q}}\psi|\partial_{{\boldsymbol\theta}_{j,l}}}-\braket{\partial_{{\boldsymbol\theta}_{i,q}}\psi|\psi}\braket{\psi|\partial_{{\boldsymbol\theta}_{j,l}}\psi}]\nonumber\\
&\approx\mathrm{Re}[
    \braket{r|U_{\boldsymbol\theta}^{\dagger,i,q} U_{\boldsymbol\theta}^{j,l}|r}
    -
    \braket{r|U_{\boldsymbol\theta}^{\dagger,i,q} U_{\boldsymbol\theta}|r}
    \braket{r|U_{\boldsymbol\theta}^{\dagger} U_{\boldsymbol\theta}^{j,l}|r}
].
\end{align}
The corresponding operator products are naturally expressed as longer time-evolution operators of the same form as Eq.~\ref{uthetaearly} with the given parameters ${\boldsymbol\theta}$ as 
\begin{equation}  
U^{\dagger,i,q}_{{\boldsymbol\theta}} U_{\boldsymbol\theta} = \hat{T}[e^{-i\int_0^2\sum_{j,k} ({\boldsymbol\theta}_{j,k}+\delta \hat{e}_{j,k}\hat{e}_{i,q}{\Theta}(t-1)) \sin(\pi k t) H_j dt}],
\end{equation} 
and analogously $U^{\dagger,i,q}_{{\boldsymbol\theta}} U_{\boldsymbol\theta}^{j,l}$ and $U^{\dagger}_{{\boldsymbol\theta}} U_{\boldsymbol\theta}^{j,l}$.
${\Theta}$ is the Heaviside-function such that the parameter ${\theta}_{i,q}$ is slightly altered by $\delta$ at $t=1$.
The Fubini-Study metric $g$ with respect to $\ket{r}=U_r\ket{0}^{\otimes n}$ can be measured by evaluating $\bra{0}^{\otimes n_\mathrm{q}}U^\dagger_rU^{\dagger,i,q}_{{\boldsymbol\theta}} U_{\boldsymbol\theta} U_r\ket{0}^{\otimes n_\mathrm{q}}$. 
Solving the linear system of Eq.~\ref{qng} yields the quantum natural gradient descent step. 
For very large experimental setups, determining the curvature with respect to only a select subset of ${\boldsymbol\theta}$ can be a beneficial compromise in terms of time-efficiency.

\end{document}